\documentclass[journal,twoside,web]{ieeecolor}
\usepackage{jsen}
\usepackage{cite}
\usepackage{amsmath,amssymb,amsfonts}
\usepackage{algorithmic}
\usepackage{graphicx}
\usepackage{textcomp}
\usepackage{wrapfig}
\usepackage{stfloats}

\def\BibTeX{{\rm B\kern-.05em{\sc i\kern-.025em b}\kern-.08em
    T\kern-.1667em\lower.7ex\hbox{E}\kern-.125emX}}
\definecolor{abstractbg}{rgb}{0.89804,0.94510,0.83137}
\begin{document}
\title{Adjustable Low-Cost Highly Sensitive Microwave Oscillator Sensor for Liquid Level Detection}
\author{Mojtaba Joodaki, \IEEEmembership{Senior Member, IEEE} and Mehrdad Jafarian
\thanks{Manuscript received ???; revised ???; accepted
???. \textit{(Corresponding author: Mojtaba Joodaki.)}}
\thanks{Mojtaba Joodaki is with the School
of Computer Science and Engineering, Constructor University, 28759
Bremen, Germany (e-mail: mjoodaki@constructor.university).}
\thanks{Mehrdad Jafarian is with the Department of Electrical Engineering, Faculty of Engineering, Ferdowsi University of Mashhad, Mashhad 9177948974, Iran (e-mail: mehrdad.jafarian.eng@gmail.com).}
}

\IEEEtitleabstractindextext{%
\fcolorbox{abstractbg}{abstractbg}{%
\begin{minipage}{\textwidth}%
\begin{wrapfigure}[20]{r}{4in}%
\includegraphics[width=3.8in]{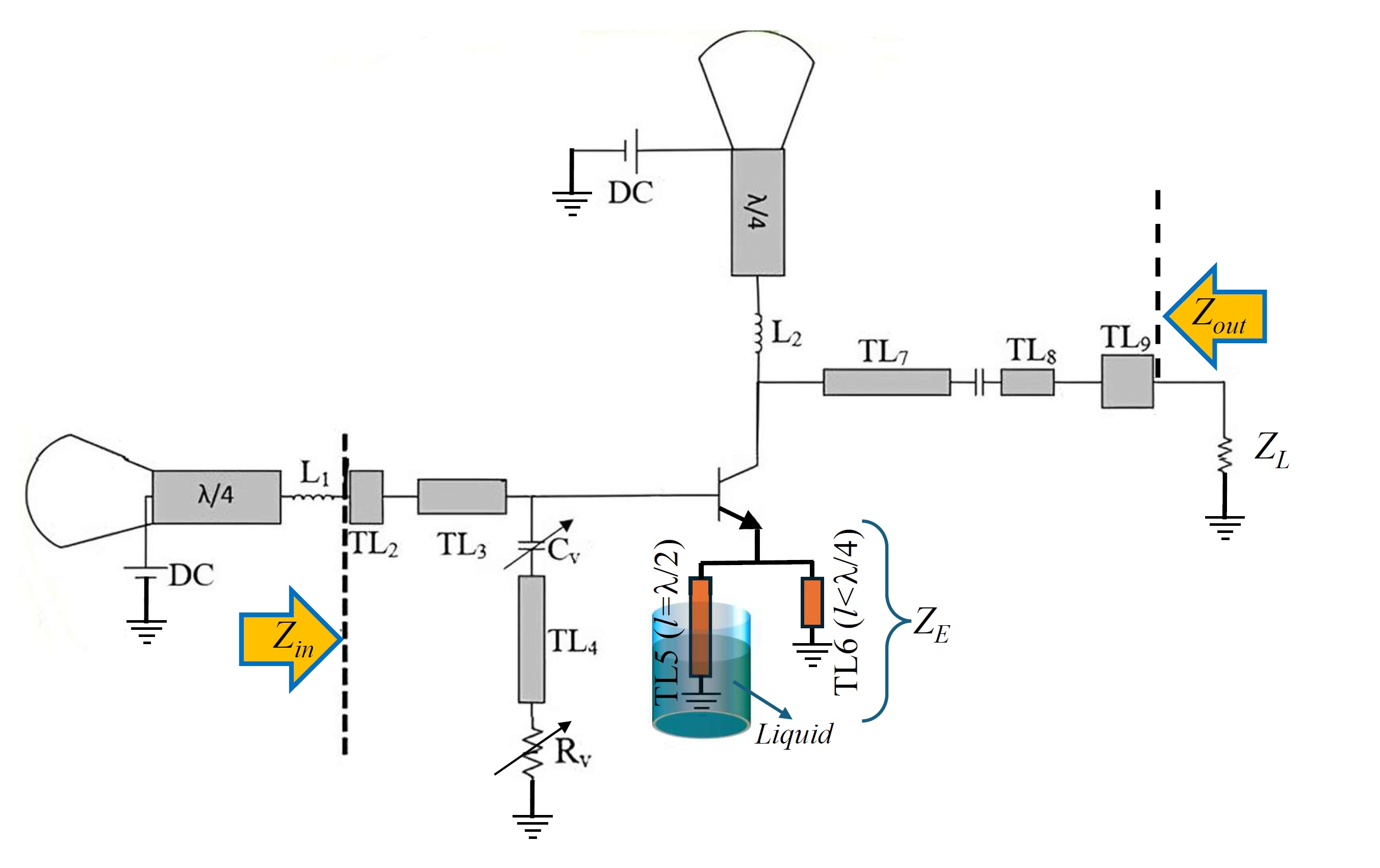}%
\end{wrapfigure}%
\begin{abstract}
This paper explores the implementation of a low-cost high-precision microwave oscillator sensor with an adjustable input resistance to enhance its limit of detection (LoD). To achieve this, we introduce a \textit{Z$_{2}$} branch in the input network, comprising a transmission line, a capacitor (\textit{C$_{B}$}) and a resistor (\textit{R$_{V}$}). The sensor is tested with eight different liquids with different dielectric constants, including water, IV fluid, milk, ethanol, acetone, petrol, olive oil, and Vaseline. By fine-tuning the \textit{Z$_{2}$} branch, a clear relationship is found between $\varepsilon_{r}$ of materials and R$_{V}$.Our experimental results demonstrate outstanding characteristics, including remarkable linearity (nonlinearity \textless 2.44\%), high accuracy with an average sensitivity of 21 kHz/$\mu$m, and an excellent limit of detection (LoD \textless 0.05 mm). The sensor also exhibits good stability across a range of liquid temperatures and shows robust and repeatable behavior. Considering the strong absorption of microwave energy in liquids with high dielectric constants, this oscillator sensor is a superior choice over capacitive sensors for such applications. We validate the performance of the oscillator sensor using water as a representative liquid. Additionally, we substantiate the sensor's improvement through both experimental results and theoretical analysis. Its advantages, including affordability, compatibility with CMOS and MEMS technologies, and ease of fabrication, make it an excellent choice for small-scale liquid detection applications.
\end{abstract}

\end{minipage}}}

\maketitle 

\section{Introduction}
\label{sec:introduction}
\IEEEPARstart{R}{ecently}, liquid level sensors have found a multitude of applications. Sensor classifications depend on several factors, including the desired accuracy, the nature of the materials to be measured, and the operational conditions. Liquid level sensors, designed to monitor the levels of various liquids, whether hazardous or benign, are categorized into small- or large-scale measurements \cite{b1}.

The key attributes sought in liquid level sensors are stability, resolution (often defined as the limit of detection, LoD) \cite{b2}, and cost. In smaller scale applications, these sensors serve a wide array of industries. They ensure precise drug dosing in medical applications and manage oil and fuel levels in the automotive sector. For larger-scale measurements, such as those in dams, petrochemical facilities, and petroleum product storage tanks, the requirements may not demand ultra-precision measurements.
One of the most prevalent sensor types for liquid level detection is the capacitive sensor. These sensors are cost-effective and exhibit commendable temperature stability \cite{b3, b4}. However, their insulation coatings can influence the measurement quality and they suffer from the parasitic capacitor around the sensor. Even in some cases, the parasitic capacitor can accumulate, leading to reduced accuracy over time \cite{b5, b5b}. Although capacitive sensors come in various shapes, such as cylindrical or interdigital, and have seen performance enhancements, their common application for water level detection with a suitable LoD remains relatively uncommon due to the high absorption of microwave energy in liquids with high dielectric constants \cite{b6}. Intriguingly, interdigital sensors, which share similarities with cylindrical capacitive sensors, are constructed with two combs on a printed circuit board (PCB). Their applications, primarily on a small scale, tend to be more costly due to their intricate structures \cite{b7, b8}.
\begin{figure*}[!h]
\centerline{\includegraphics[width=5in]{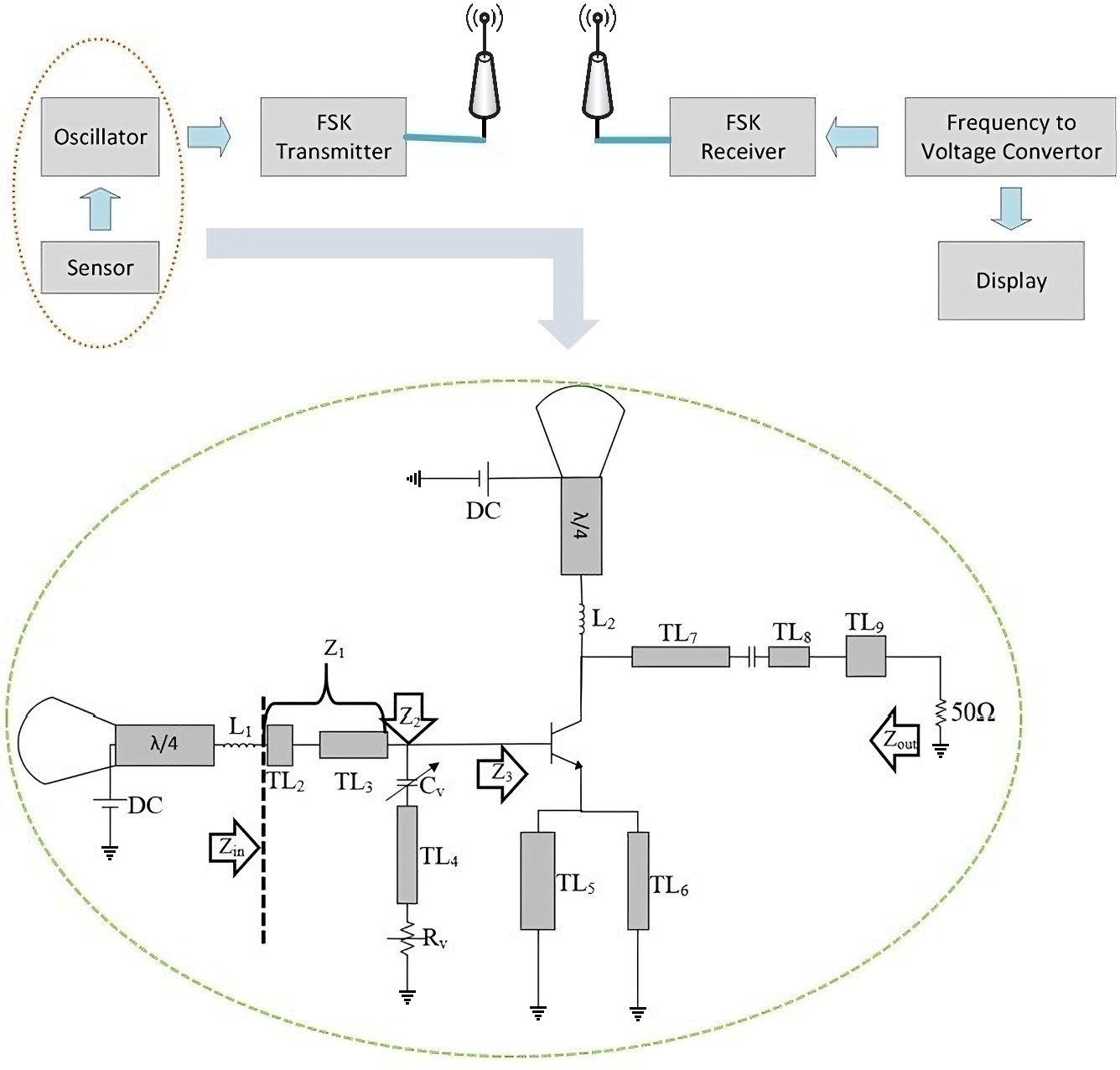}}
\caption{(a)Combining the oscillator sensor with a FSK transmitter/receiver system \cite{b16}. (b)Topology of the input and output networks to build the oscillator sensor.}
\label{fig1}
\end{figure*}
Other techniques, like quartz crystal microbalance (QCM) technology, have been used for small-scale liquid detection with high resolution. However, these sensors require costly material coatings and exhibit suboptimal performance at high frequencies with limited bandwidth. As a result, they are less suited for applications in emerging technologies like IoT and 5G networks \cite{b9, b10}.
Optical sensing technologies, including fiber Bragg grating (FBG) and tilted fiber grating (TFG), have demonstrated efficacy in measuring liquid levels for both small and large-scale applications. Although their performance shines brightest in small-scale applications while offering outstanding linearity and exceptional accuracy in stable environments, their high sensitivity to temperature and relatively high cost impede their broader application \cite{b11, b12, b13}.

On the other hand, microwave sensors are increasingly being integrated into 5G/6G and IoT ecosystems, enabling real-time monitoring and data collection for various applications \cite{b14a, b14b, b14c, b14d, b14e, b14f}. Microwave planar sensors offer impressive resolution, cost-effectiveness, compatibility with CMOS and MEMS technologies, and ease of fabrication, making them ideal candidates for liquid level detection. Over the last two decades, most microwave planar sensors have been built around passive resonators. However, these designs often suffer from low-quality factors, which reduce sensitivity and degrade LoD \cite{b2}. This emphasizes the need for active sensors that can enhance resolution and reduce LoD. To address this need, this work presents a novel microwave oscillator sensor designed to provide high-resolution and stable liquid level detection. Furthermore, in an era where seamless connectivity to 5G networks is imperative, the sensor's ability to integrate into a network of interconnected sensors becomes paramount. Analyzing data from such a network allows for precise determinations of the liquid level \cite{b15}. A schematic representation of a wireless sensor network (WSN) is depicted in Fig. 1. The sensor's oscillatory nature also plays a crucial role in facilitating easy communication within the WSN. These sensors produce stable output signals that can be easily used in frequency-shift keying (FSK) transmitters \cite{b16}.
This article introduces an innovative oscillator sensor designed to address the demand for high sensitivity, cost-effectiveness, and compact size. This new oscillator sensor not only offers exceptional stability to temperature fluctuations, but also offers the flexibility to adjust its frequency, making it an ideal choice for a wide range of applications.

Section II of this article covers the sensor design and analysis and it explores the \textit{Z$_{2}$} branch’s impact on the circuit’s input resistance and its effectiveness in attenuating the parasitic harmonics generated at the output. This analysis will shed light on the critical technical aspects of sensor operation.
In Section III, the sensor fabrication and measurement setup are explained. Practical results obtained from our experimental work, providing valuable insights into the real-world performance of the sensor and its suitability for various applications, are presented in Section IV.
Finally, in Section V, we draw together the threads of our research and provide conclusive insights and recommendations. By exploring the potential of this novel oscillator sensor, we aim to contribute to the field of sensor technology, offering a solution that not only addresses current challenges, but also lays the groundwork for future advancements.

\section{Sensor Design and Analysis}
The circuit presented in Fig. 1 is an oscillator sensor with negative resistance, designed for the precise detection of liquid levels. It accommodates liquids with varying dielectric constants. The strength of this design lies in its adaptability; by changing components in the input network, the oscillation frequency can be finely tuned. In this configuration, the oscillator reliably operates at a frequency of 3.6 GHz.
To enhance its functionality, two parallel transmission lines have been incorporated into the transistor emitter, both shorted at their respective ends. One of these transmission lines serves as a short-circuited $\lambda$/2 resonator dedicated to liquid level measurements. All the geometric parameters of this resonator have been meticulously adjusted to ensure that it resonates at 3.6 GHz. Therefore, when this resonator is connected to the oscillator, it satisfies the Barkhausen criteria for the desired oscillation frequency. Another transmission line with a length shorter than $\lambda$/4 behaves as an inductor to ensure proper functioning of the oscillator under any condition, for example, even when the sensing transmission line is in the air.
This oscillator sensor is a versatile tool for detecting liquid levels, and its adaptability to various dielectric constants makes it an excellent choice for a wide range of applications. The careful design of the resonator ensures that it harmonizes with the oscillator, resulting in accurate and reliable liquid level measurements.

\subsection{Circuit's Input Resistance Analysis}

The analysis of input resistance is a crucial aspect of level detection sensors. These sensors need to be adaptable to different materials in order to achieve optimal performance. Consequently, it is essential to employ a method that allows the calibration of these sensors across a range of materials with varying dielectric constants. This approach offers some advantages: firstly, it is cost-effective, as it eliminates the need to purchase multiple sensors for different materials, and secondly, it facilitates the calibration of microwave sensors, whether they operate through contact or non-contact methods, in alignment with the specific environmental and measurement conditions.
In the realm of oscillator design, two key criteria, frequency stability and phase noise, take center stage. These factors are significantly influenced by the input negative resistance at the start of the oscillation. Therefore, it is imperative to perform a comprehensive examination to quantify the effects of various factors on the negative resistance perceived from the input. Controlling the negative resistance observed from the input network is a straightforward approach to manage the various non-linear parameters within an oscillator. The optimal range for negative resistance typically falls between -50 and -5 ohms \cite{b17}. An excessive negative resistance is undesirable because its absorption requires a large nonlinear signal level, which degrades noise performance by increasing higher harmonics that may cause spurious modes. Therefore, moderate negative resistance values are typically employed to achieve rapid oscillator response and lower harmonic distortions. 

\begin{figure}[!t]
\centerline{\includegraphics[width=2.5 in]{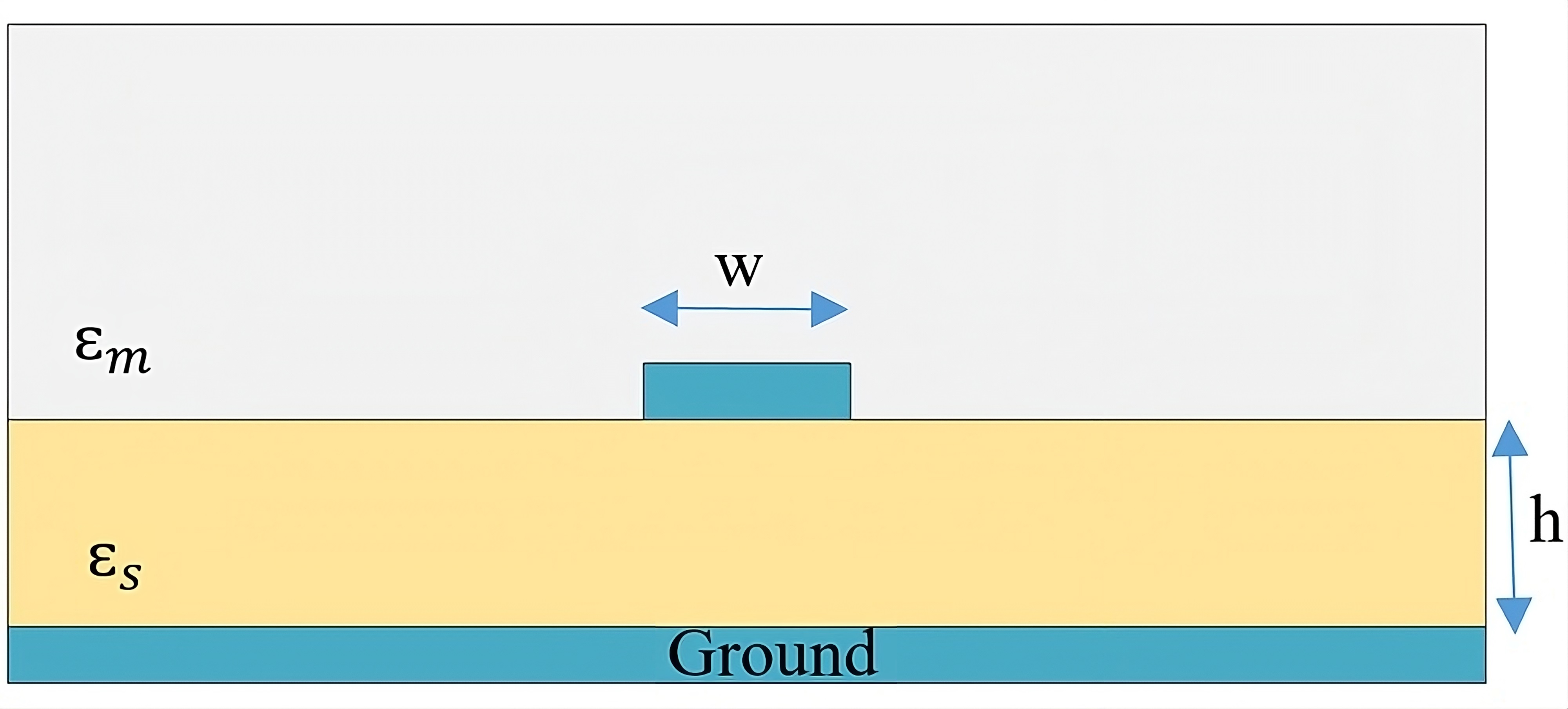}}
\caption{Geometry of the microstrip transmission line (MSTL) used.}
\label{fig1}
\end{figure}
It should be noted that the effective dielectric constant of the microstrip transmission line (MSTL) is subject to variation due to changes in fluid level, according to the following equation \cite{b18}:

\begin{equation}
    \varepsilon_e = \frac{\varepsilon_s + \varepsilon_m}{2} + \frac{\varepsilon_s - \varepsilon_m}{2(1 + 12h/w)}
\end{equation}
where $\varepsilon_{s}$, $\varepsilon_{m}$, \textit{w} and \textit{h} are substrate dielectric constant, liquid dielectric constant, signal trace width and substrate thickness, respectively. Then, the characteristic impedance of the transmission line, which depends on the dimensions of the transmission line, can be determined based on the following equations \cite{b19}:
\begin{equation}
    Z_0 = \frac{60}{\sqrt{\varepsilon_e}} \ln\left( \frac{8h}{w} + \frac{w}{4h} \right)
\end{equation}

\begin{equation}
    Z_0 = \frac{120\pi}{\sqrt{\varepsilon_e}} \left[ 1.393 + 0.667 \ln\left(\frac{w}{h} + 1.444\right) \right]
\end{equation}
On the other hand, a short-circuited $\lambda$/2 transmission line resonator can be modeled using the following approximated lumped elements \cite{b20}:
\begin{equation}
    R_{eq} \text{(resistance)} = Z_0 \alpha \ell \quad (\Omega)
\end{equation}

\begin{equation}
    L_{eq} \text{(inductance)} = \frac{Z_0\pi}{2\omega} \quad (H)
\end{equation}

\begin{equation}
    C_{eq} \text{(capacitance)} = \frac{1}{L_{eq}\omega_0^2} \quad (F)
\end{equation}
where $\alpha$, $\ell$ and $\omega_0$ are loss of the line, length of line and resonance frequency, respectively.
These equations demonstrate how variations in the dielectric constant affect the electrical characteristics of the components within the resonator. Based on equations (1) to (6), any increase in $\varepsilon_{e}$ (the effective permittivity) of materials results in the following effects on the equivalent lumped elements:
\begin{equation}
\uparrow \varepsilon_{e} =
\left\{
\begin{array}{c}
\uparrow C_{\text{eq}} \\
\downarrow R_{\text{eq}} \\
\downarrow L_{\text{eq}}
\end{array}
\right.
\end{equation}

The equivalent circuit of the oscillator with lumped elements is illustrated in Fig. 3(a). The equivalent circuit of the transistor is made up of \textit{L$_{B}$} (base parasitic inductance), \textit{r$_{bb}$} (base resistance), \textit{r$_{be}$} (base-emitter resistance or \textit{r$_{\pi}$)}, \textit{C$_{be}$} (base-emitter capacitance) and \textit{L$_{e}$} (emitter parasitic inductance), and \textit{C$_{cb}$} (collector-base capacitance). If the lumped elements responsible for frequency tuning in the circuit, specifically \textit{R$_{V}$} (resistor) and \textit{C$_{B}$} (capacitor), are placed within the \textit{Z$_{2}$} branch, they exert the most substantial influence on the negative input resistance of the circuit. However, incorporating these elements in this branch results in rapid alterations in two fundamental parameters that determine the oscillation conditions in the oscillators, as can be seen in Fig. 6; \(|S_{11}|\) (magnitude of the scattering parameter S$_{11}$) and $\angle S_{11}$ (phase of the scattering parameter S$_{11}$). This means that the oscillator frequency experiences significant variations. As a consequence, this can limit the dynamic range (DR) of the sensor. Therefore, after the appropriate design of the oscillation at the relevant frequency, this branch is fixed. In the following sections, we will analyze how the variations in \textit{R$_{V}$} and \textit{C$_{B}$} affect the observed negative input resistance at the input of the transistor.

\begin{figure*}[!t]
\centerline{\includegraphics[width=5.5 in]{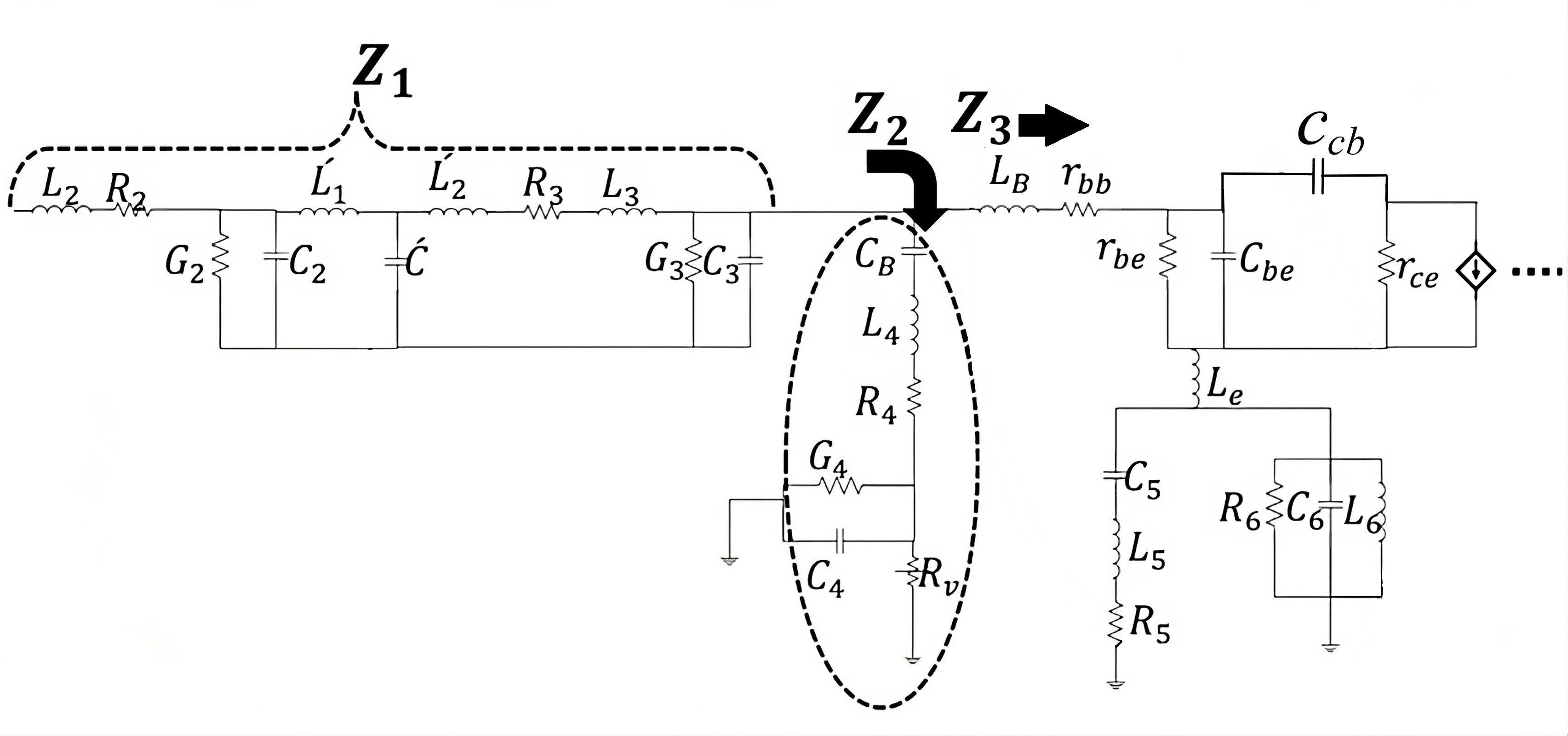}}
\[{(a)}\]
\centerline{\includegraphics[width=5.5 in]{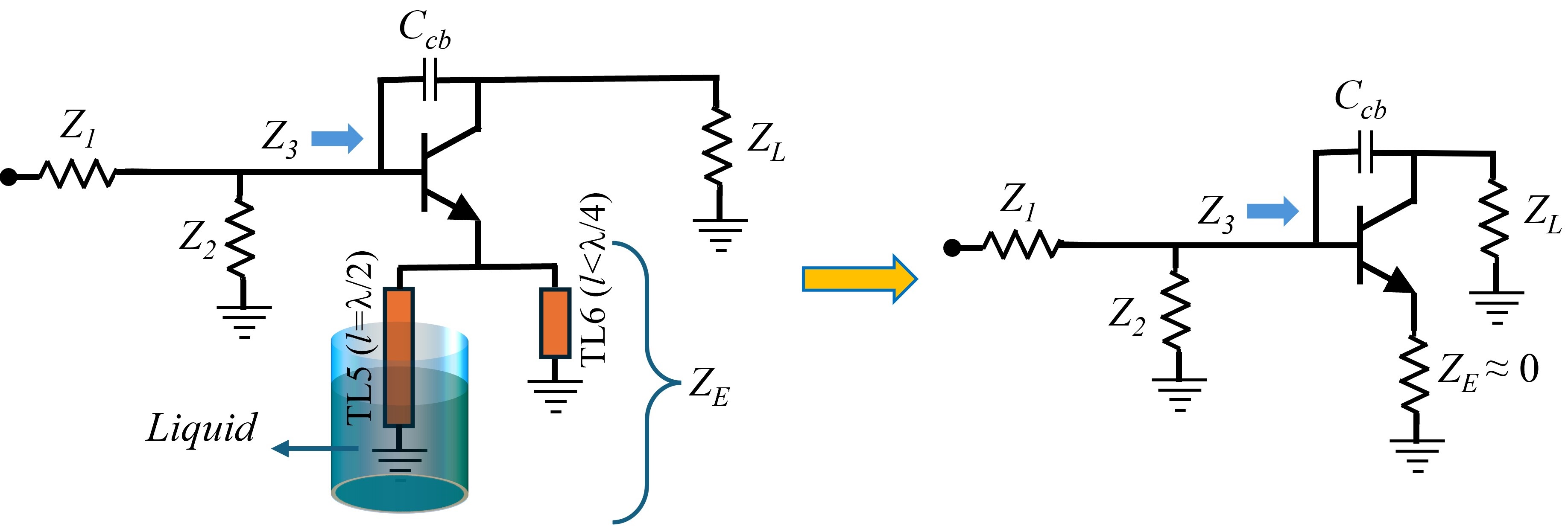}}
\[{(b) \hspace{8cm} (c)}\]
\caption{(a) Small-signal equivalent circuit, (b) ac schematic and (c) simplified ac circuit of the oscillator sensor.}
\label{fig1}
\end{figure*}

Fig. 3(b) shows a simplified schematic of the oscillator circuit. Since the transmission line TL5, which acts as the sensor probe inside the liquid, is resonating at its series resonance frequency, the emitter impedance ($Z_E$) is very small, and the circuit will be reduced to Fig. 3(c). Using Fig. 3(c) admitance seen at the base of the transitor is equal to: 

\begin{equation}
Y_{3} = \frac{1}{Z_{3}} = \frac{1 + g_{m} Z_{L}}{Z_{L} - \frac{j}{\omega C_{cb}}}
\end{equation}
in which $g_m$ is the transistor transconductance and $Z_L$ is the impedance seen at the collector. As $C_{cb}$ is very small, $Y_3$ is reduced to: 
\begin{equation}
Y_{3} \approx j \omega C_{cb} \left( 1 + g_{m} Z_{L} \right)
= -\omega C_{cb} g_{m} X_{L} + j \omega C_{cb}
\end{equation}
in which $Z_L=jX_L$. The real part of $Y_3$ will provide the necessary negative resistance for the oscillation, while it can be adjusted by $R_V$ in the $Z_2$ branch.

\subsection{Effect of Z$_{2}$ branch on the output signal phase noise and distortion}
In this study, we investigate two distinct categories of materials to optimize the sensor oscillator for liquid level measurement. The first category consists of materials with dielectric constants less than 25, which are often prone to flammability, and the second category encompasses materials with high dielectric constants. 

Initially, a 3D full-wave electromagnetic simulator, CST Studio Suite, was used to accurately design the transmission line resonator. The corresponding 3D model is illustrated in Fig. 4. The simulated S-parameters were then saved in a touchstone file and imported into ADS software for nonlinear harmonic balance simulation and further analysis. Simulations were performed and the results were validated experimentally. The sensor is specifically designed to measure the liquid levels of substances with different dielectric constants. To optimize sensor stability and minimize phase noise for a specific liquid, the sensor is immersed 3.5 cm in the liquid, and the \textit{R$_{V}$} is fine-tuned for optimal performance. For example, for water, we changed \textit{R$_{V}$} from 200 $\Omega$  to 50 $\Omega$, while keeping the capacitance (\textit{C$_{B}$}) constant at 100 nF. This resulted in a reduction in unwanted harmonics and a noticeable improvement in the quality of the output signal. The calculated results using the nonlinear harmonic balance simulator in ADS software of Keysight are illustrated in Fig. 5, where a 35 dB reduction in the second harmonic is observed, signifying a strong improvement in the quality of the output signals. The optimal \textit{R$_{V}$} of 82 $\Omega$ is selected to have a strong output signal while minimizing undesired harmonics.  

\begin{figure}[!b]
\centerline{\includegraphics[width=\columnwidth]{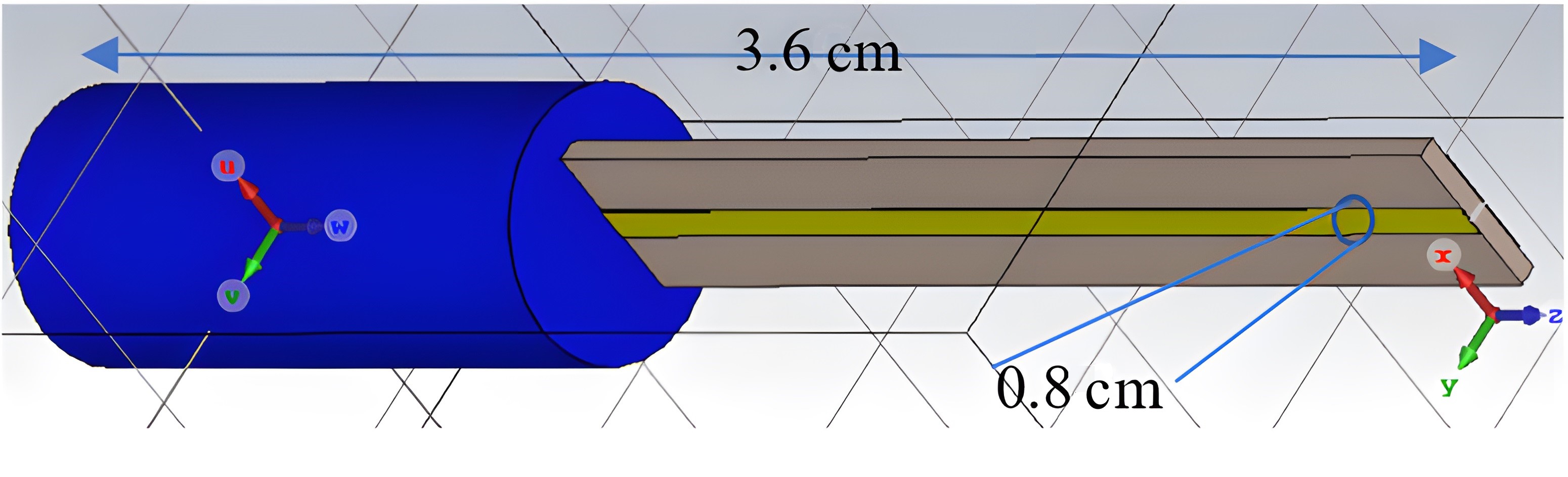}}
\caption{3D full-wave electromagnetic model of the sensor resonator in CST Studio Suite and detailed specifications of its dimensions.}
\label{fig1}
\end{figure}

\begin{figure}[!b]
\centerline{\includegraphics[width=3 in]{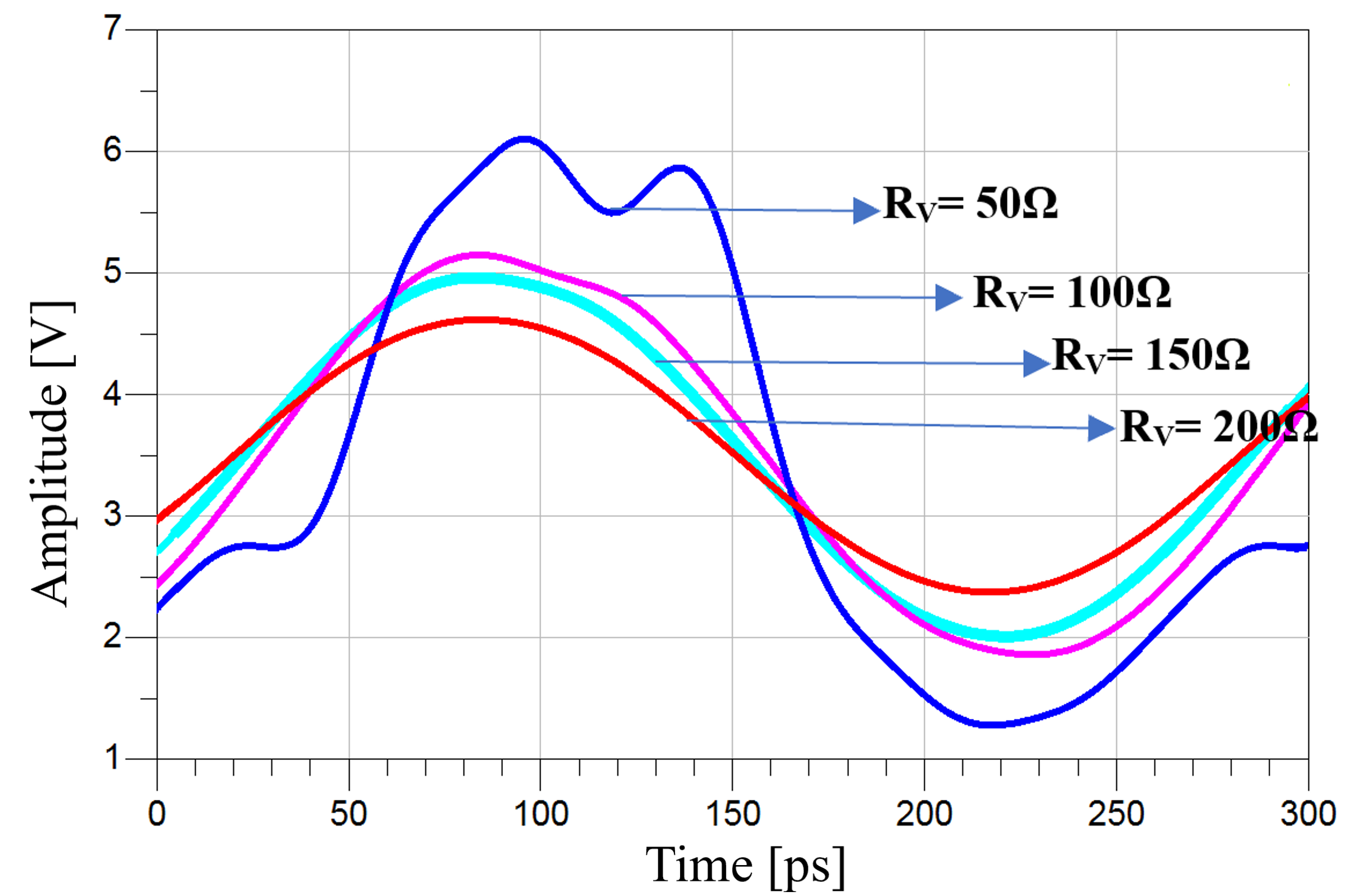}}
\[{(a)}\]
\centerline{\includegraphics[width=3 in]{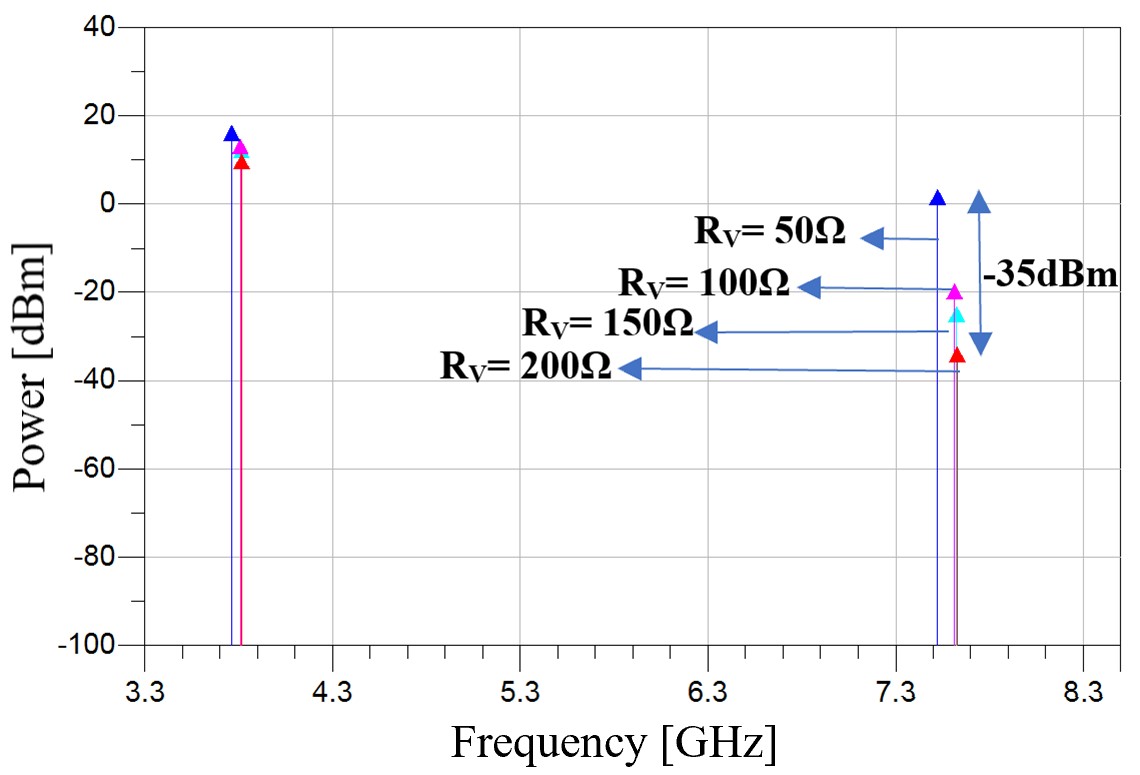}} 
\[{(b)}\]
\caption{Simulation results using harmonic balance simulator in ADS of Keysight by changing \textit{R$_{V}$} from 200 $\Omega$  to 50 $\Omega$:(a) time domain and (b) frequency domain.}
\label{fig1}
\end{figure}

The optimal \textit{R$_{V}$} values required for minimum harmonic distortion for different liquids were determined theoretically and experimentally. The results are presented in Fig. 6 and summarized in Table I. A relatively linear inverse relationship was observed between the dielectric constant and the optimal \textit{R$_{V}$} value. It should be noted that liquids with dielectric constants between 30 and 60 were not available for measurement, so the \textit{R$_{V}$} values in that range were extrapolated from other measured results. However, as you see in the following section, for liquids with higher dielectric constants we have performed various experiments, because this sensor offers higher resolution (LoD) and good sensitivity for such liquids.

\begin{figure}[!t]
\centerline{\includegraphics[width=2.5 in]{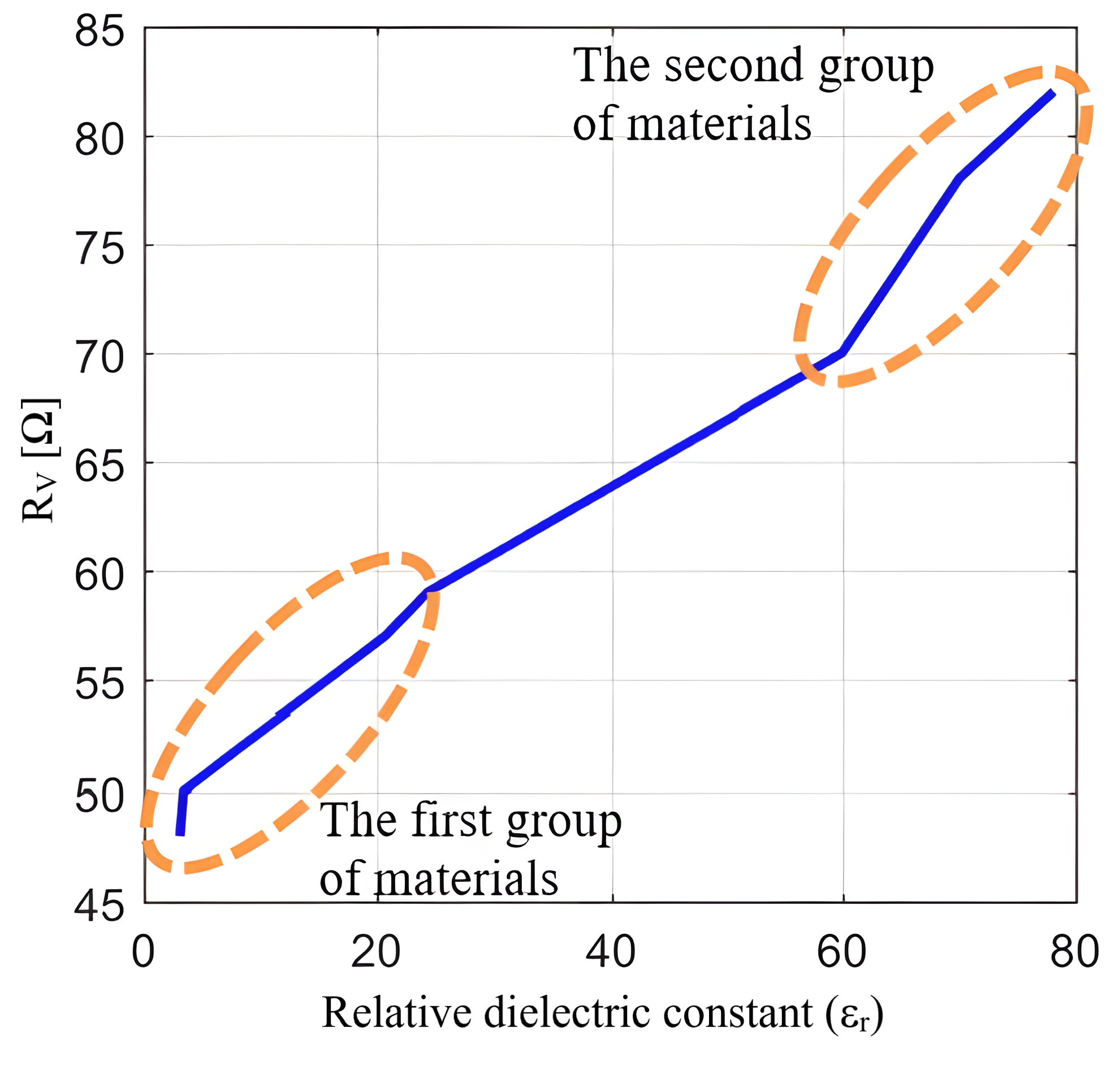}}
\caption{Optimum \textit{R$_{V}$} resistances obtained through experiments and simulations versus the dielectric constant of the measured liquids.}
\label{fig1}
\end{figure}

In addition, we experimentally measured the phase noise using the optimal values of \textit{R$_{V}$} obtained above. The measured signal spectrum of the circuit at a temperature of 25 °C is presented in Fig. 7. To measure the sensor output signal, an NS-132 spectrum analyzer of NEX1 Future was used. According to the equipment user manual, the following equation should be used to correct phase noise measurement errors caused by the Gaussian filter and the IF amplifier.
$L_{\text{Corr}}(f) =$\\
\begin{equation}
L_{\text{meas}}(f) - 10\log(\text{RBW}) - 0.79 + 2.5 \qquad [dBc/Hz]
\end{equation}

where \textit{L(f)$_{Meas}$} is the measured phase noise, \textit{L(f)$_{Corr}$} is the corrected final phase noise, and RBW is the resolution bandwidth. Key factors involved in these corrections include the presence of a resolution bandwidth (RBW) of the Gaussian filter with a bandwidth 1.2 times the final value and the 2.5 dB error induced by the IF amplifier in this device which should be considered in correcting the result. The outcomes of both practical and theoretical evaluations are described in Table I. Fig. 7 shows the measured phase noise for a 100 kHz offset frequency. As the measured phase noise (\textit{L(f)$_{Meas}$}) before correction is -73.7 dB for the offset frequency of 100 kHz, then using (10) and RBW of 1 kHz the corrected phase noise is -102 dBc/Hz. 

\begin{figure}[!t]

\centerline{\includegraphics[width=3 in]{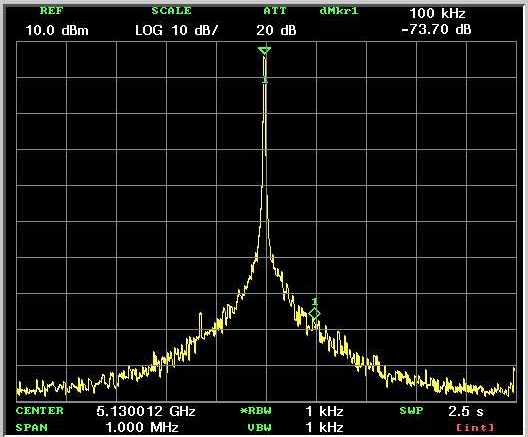}}

\caption{Measured frequency spectrum of the oscillator sensor with the optimum \textit{R$_{V}$}. The optimum \textit{R$_{V}$} is given in Table I for each measurand.}
\label{fig1}
\end{figure}

\subsection{Oscillation Analysis and Simulation Results}

Fig. 8 illustrates the 3D full-wave sensor models at various water levels in the CST Studio Suite. As already mentioned, the corresponding simulated S-parameters were then transferred to ADS software for nonlinear harmonic balance circuit simulation. Each case is represented by a specific color, and its corresponding simulated results can be seen in Fig. 9. Moreover, Figs. 9(a) and 9(b) represent the phase and magnitude criteria for oscillation (Barkhausen criteria), respectively. They indicate that the phase and magnitude criteria for stable oscillation are met as the liquid level changes in 0.5 cm increments (from 0.5 to 3.5 cm, the measurement at 1 cm is not included). Fig. 9(c) illustrates the ability of the sensor to maintain its input resistance within the range of -50 $\Omega$ to -15 $\Omega$ as the liquid level changes. As was already explained, ensuring that the input resistance remains within this range is vital for the oscillator's ability to operate in its stable oscillating condition with a low phase noise.
With these findings, it is evident that the water level sensor, operating within the specified parameters, is capable of meeting the Barkhausen criteria for stable oscillation and maintaining the desired input resistance, making it a reliable tool for measuring water levels with adaptability to varying conditions. We conducted similar investigations on different liquids with different dielectric constants. For this purpose, the corresponding optimum \textit{R$_V$} values provided in Table I should be used to enhance the performance of the oscillator sensor.

\begin{figure}[!t]
\centerline{\includegraphics[width=2.5 in]{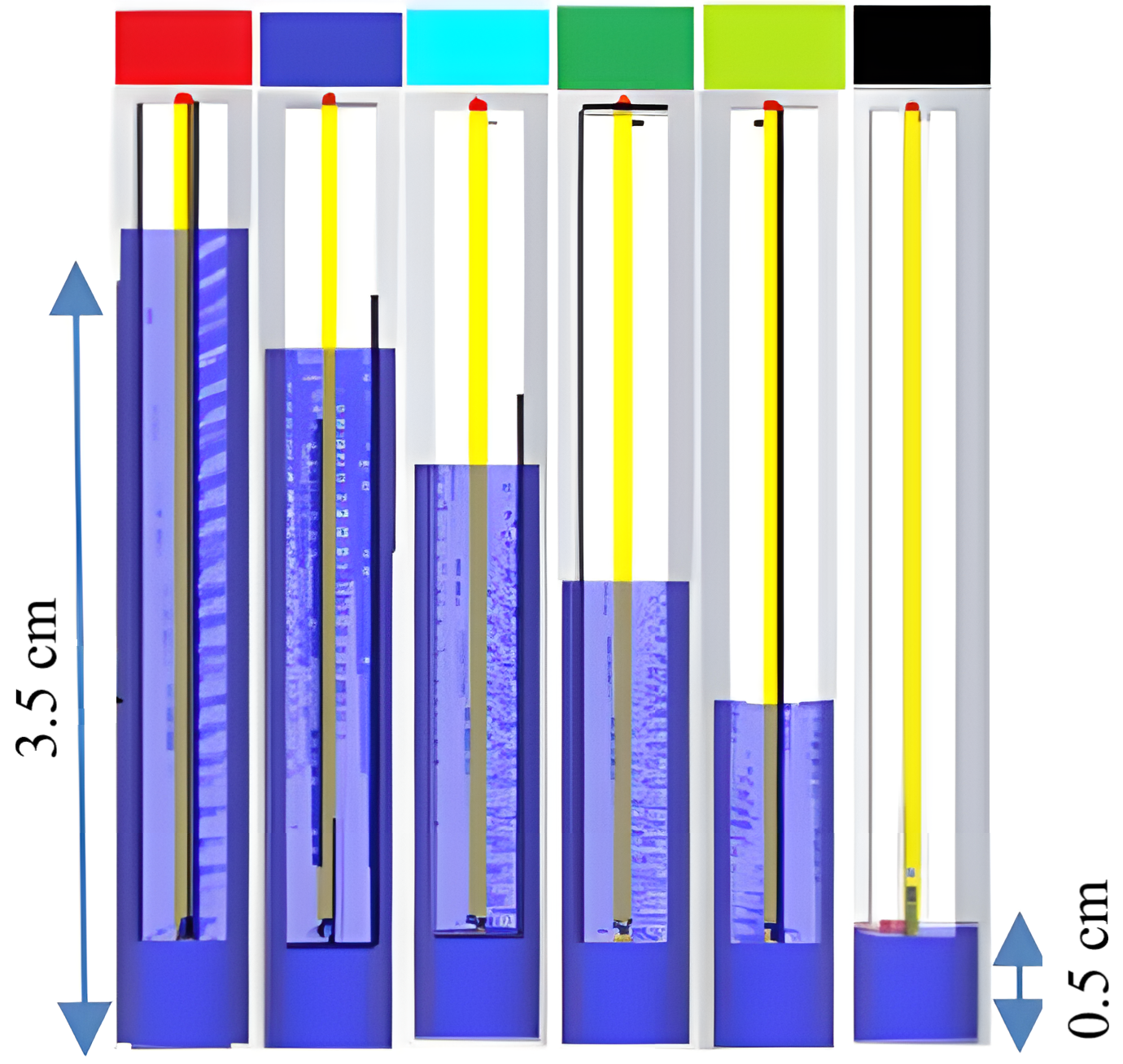}}
\caption{3D sensor models at different liquid levels. Each liquid level is indicated by a specific color.}
\label{fig1}
\end{figure}

\begin{figure*}[h]
\centering
\includegraphics[width=\textwidth]{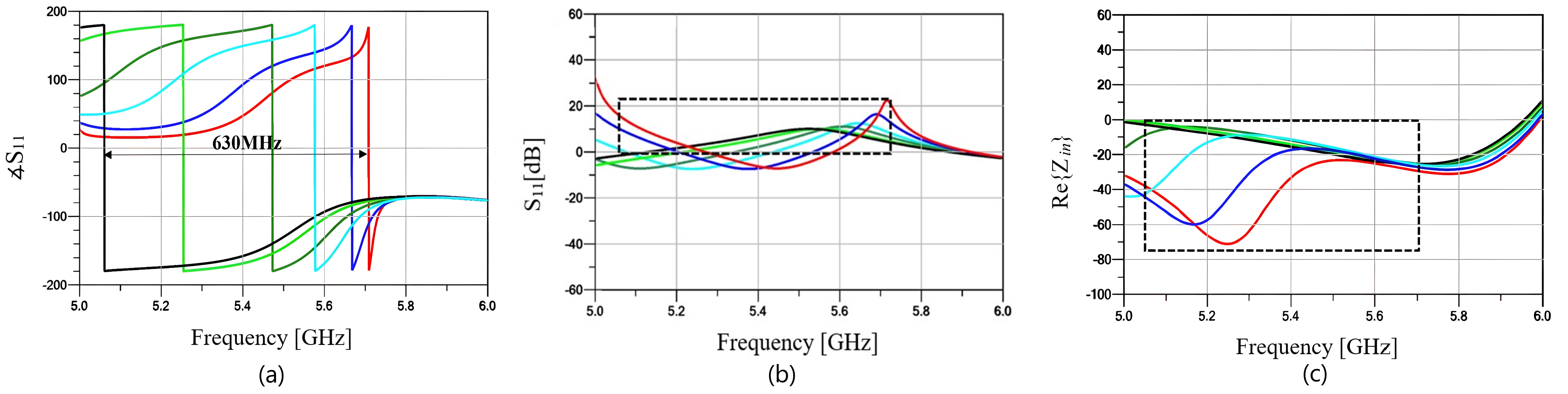}
\caption{Sensor simulation results when measuring water level by setting \textit{R$_V$} = 82 and \textit{C$_B$} = 100 pF. (a) $\angle S_{11}$, (b) $|S_{11}|$ and (c) $\text{Re} \{Z_{\text{in}}\}$ versus Frequency. $Z_{\text{in}}$is the imedance seen at base of the transistor. Each liquid level is indicated by a specific color, see Fig. 8.}
\label{fig1}
\end{figure*}

\begin{table}
\caption{Specifications of the tested materials along with the optimum R$_{V}$ values obtained theoretically and experimentally at a temperature of 25 °C.}
\label{table}
\label{tab1}
\centerline{\includegraphics[width=\columnwidth]{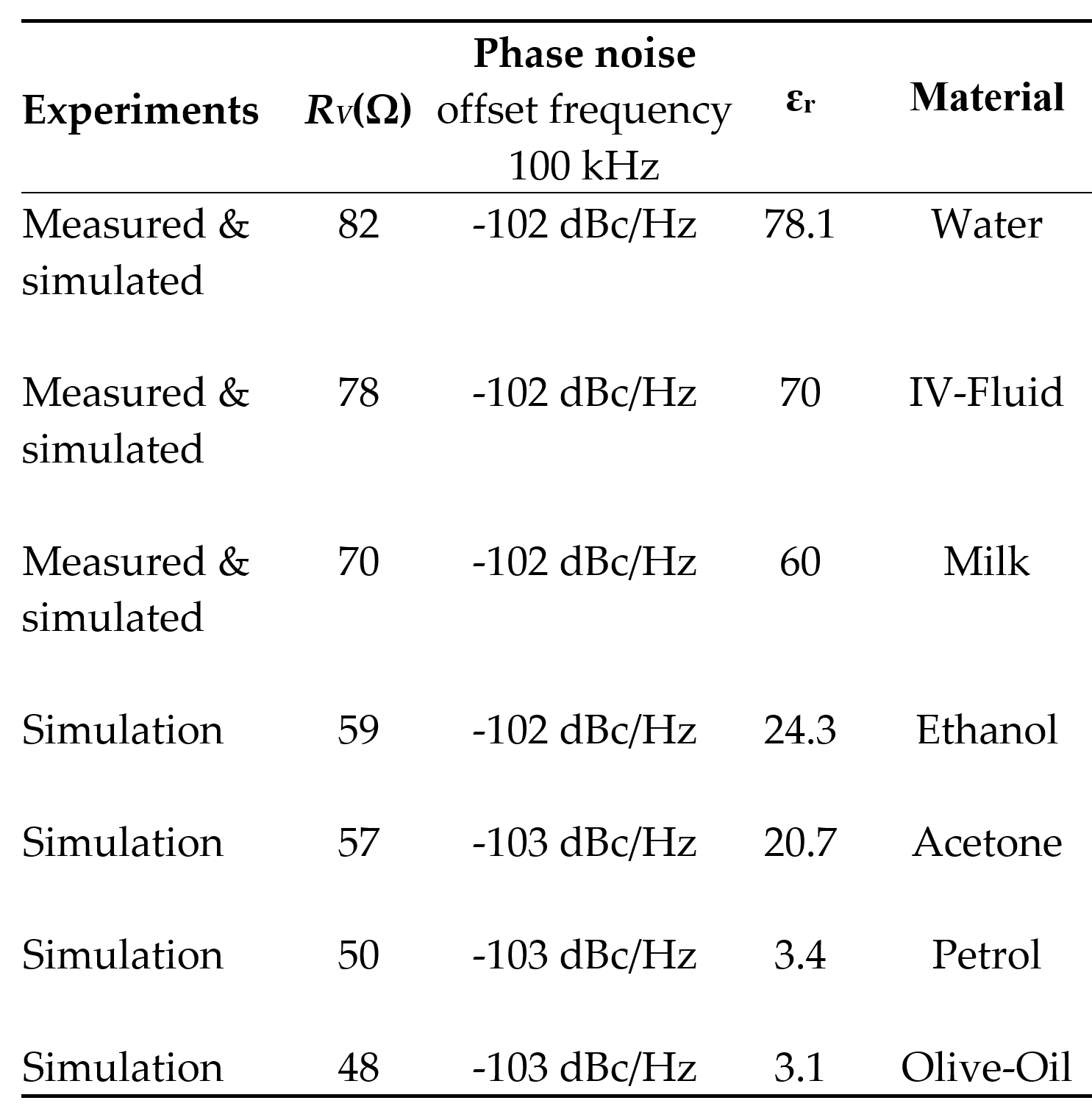}}
\end{table}

 \section{Sensor Fabrication and Measurement Setup}
The oscillator sensor circuit is fabricated on a Rogers PCB of RO4003 with dimensions of 2.6 cm by 9.6 cm using surface mounted devices (SMDs). A low-cost, highly linear, low-noise silicon NPN bipolar transistor of BFP650 used as the active component.
The transmission line resonator follows the following specifications: a height (h) of 508 µm, a width (w) of 1 mm, a relative permittivity ($\epsilon_r$) of 3.55, and a length (\textit{L}) of 70 mm. Experimental validation of sensor performance was conducted using a NEX1 Future NS-132 spectrum analyzer to assess liquid levels with different dielectric constants. Fig. 10 illustrates the oscillator sensor's measurement setup.

The water level measurements are performed at temperatures of 2 °C, 25 °C, 50 °C, and 75 °C and the results are shown in Fig. 11. The water level is changed from 0.5 cm to 3.5 cm. The results show a very linear sensor behavior. The high sensitivity is remarkable, as a 3 cm change in the water surface corresponds to a frequency change of 630 MHz at a temperature of 25 °C, indicating a very high sensitivity of this oscillator sensor. Also, the results indicate the repeatability of the experiments at different temperatures.

\begin{figure}[!h]
\centerline{\includegraphics[width=\columnwidth]{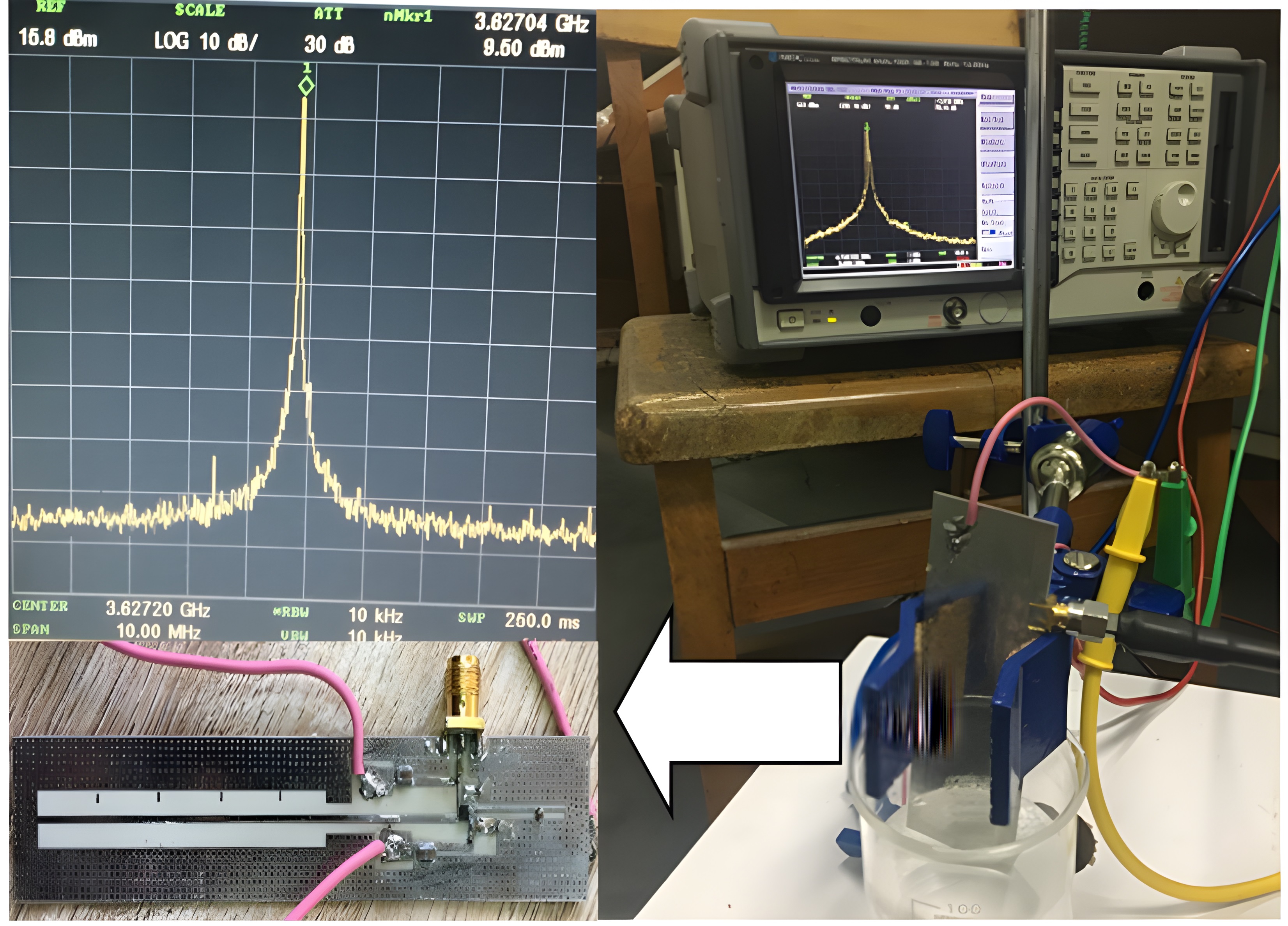}}
\caption{Measurement setup used for performing the tests on the oscillator sensor.}
\label{fig1}
\end{figure}

 \section{Measured results}
The nonlinear error quantifies how the sensor's response deviates from ideal linear behavior. To calculate the relative maximum nonlinearity error (RMNLE) of the sensor, we first need to determine the maximum deviation in the diagram from the ideal straight line and then divide this by the entire measurement range. The relative average nonlinearity error (RANLE) is another important parameter of the sensor, and is equal to the average error of all measurement points divided by the measurement range. RMNLE and RANLE of the sensor at different temperatures are outlined in Table II.
As demonstrated in Table II, the RMNLE increases with the temperature of the water, but remains below 2.7$\%$ even at 75 °C. This analysis provides crucial insights into the sensor's performance characteristics, particularly its ability to maintain accuracy across a range of liquid temperatures.

The sensitivity of the level sensor at 25 °C is calculated as follows:
\begin{equation}
\frac{\Delta f}{\Delta h} = \frac{630 \text{ MHz}}{3 \text{ cm}} = 0.021 \left( \frac{\text{MHz}}{\mu\text{m}} \right)
\end{equation}
\begin{figure*}[!h]
\centerline{\includegraphics[width=6 in]{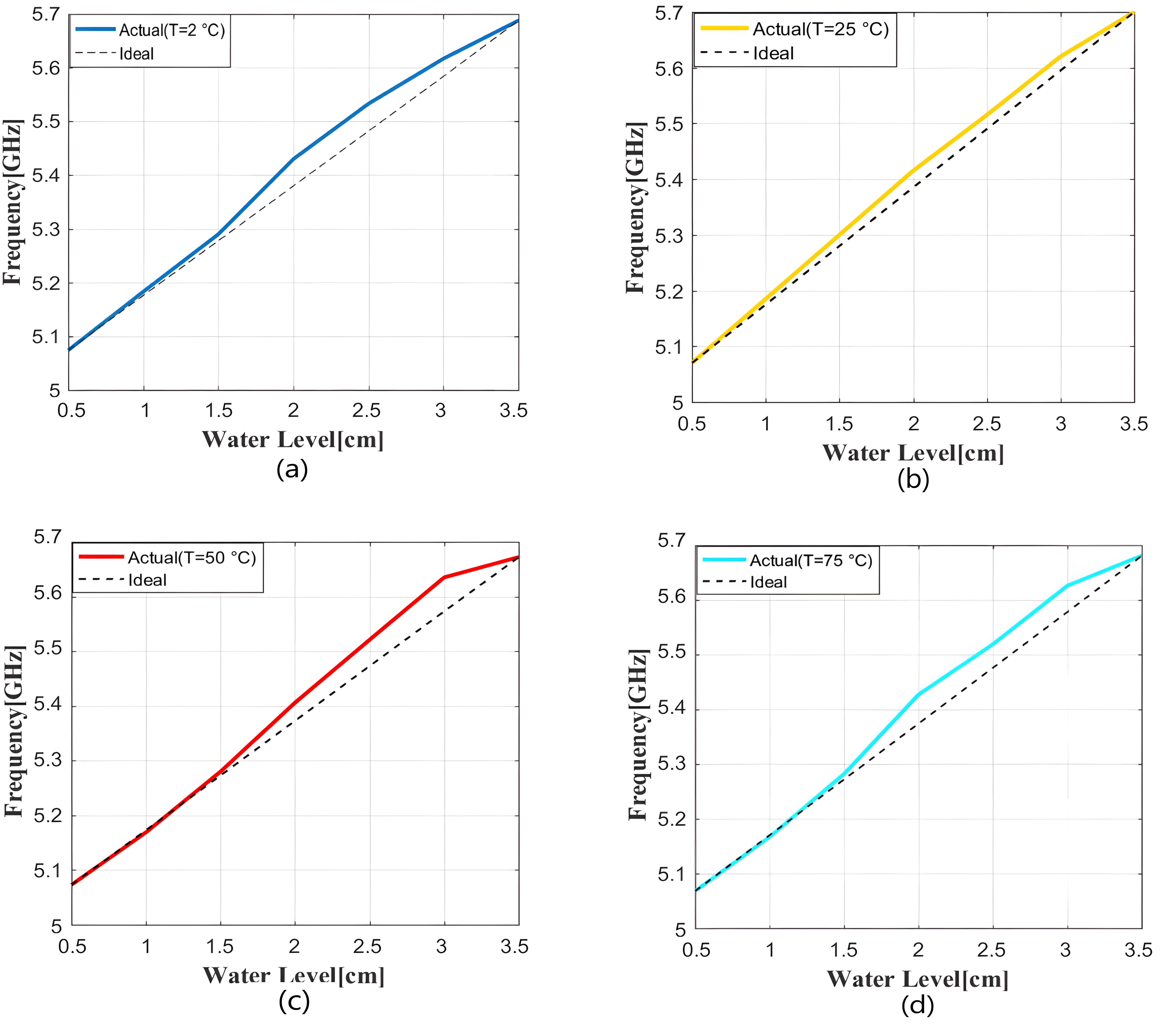}}
\caption{The measured oscillator output frequency for different water levels performed at different water temperatures: (a) T=2 °C, (b) T=25 °C, (c) T=50 °C, and (d) T=75 °C.}
\label{fig1}
\end{figure*}

\begin{table}
\caption{Variation of nonlinearity errors versus water temperature.}
\label{table}
\label{tab1}
\centerline{\includegraphics[width=3 in]{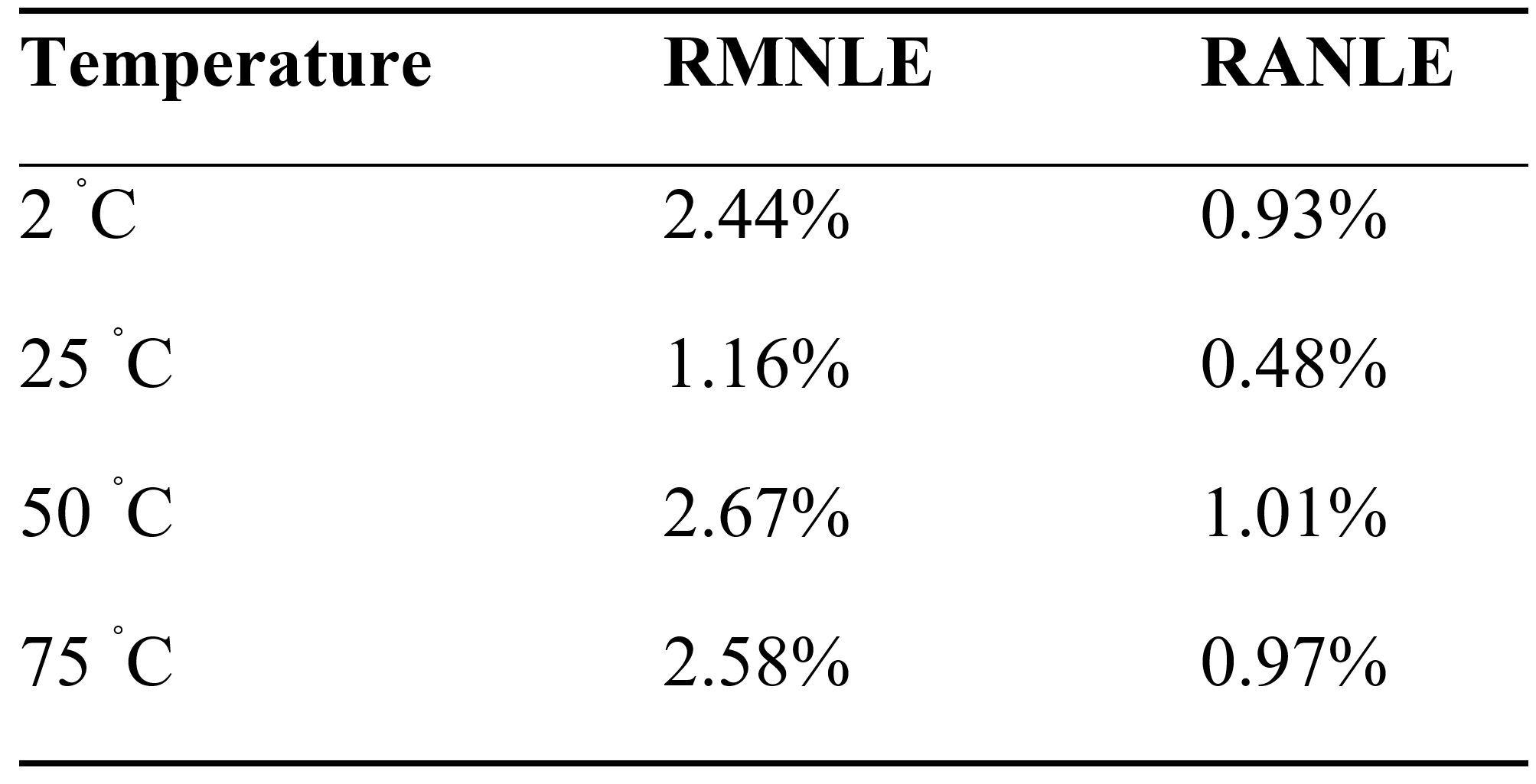}}
\end{table}
Since directly measuring LoD in our measurement setup was challenging and could introduce substantial error, the liquid was added using a vessel at least 100 times smaller than that employed in the measurement setup. This method enabled precise control of the liquid level, allowing adjustments in 0.05 mm increments. However, the result only indicates that the true LoD is less than 0.05 mm.

To calculate the relative hysteresis error, one can refer to Fig. 12 and the following equation:
\begin{equation}
\text{Relative hysteresis error} = 
\left| \frac{(Y_{\text{mn}} - Y_{\text{mp}})}{(Y_{\text{max}} - Y_{\text{min}})} \right| \times 100\%
\end{equation}

\begin{figure}[!h]
\centerline{\includegraphics[width=3 in]{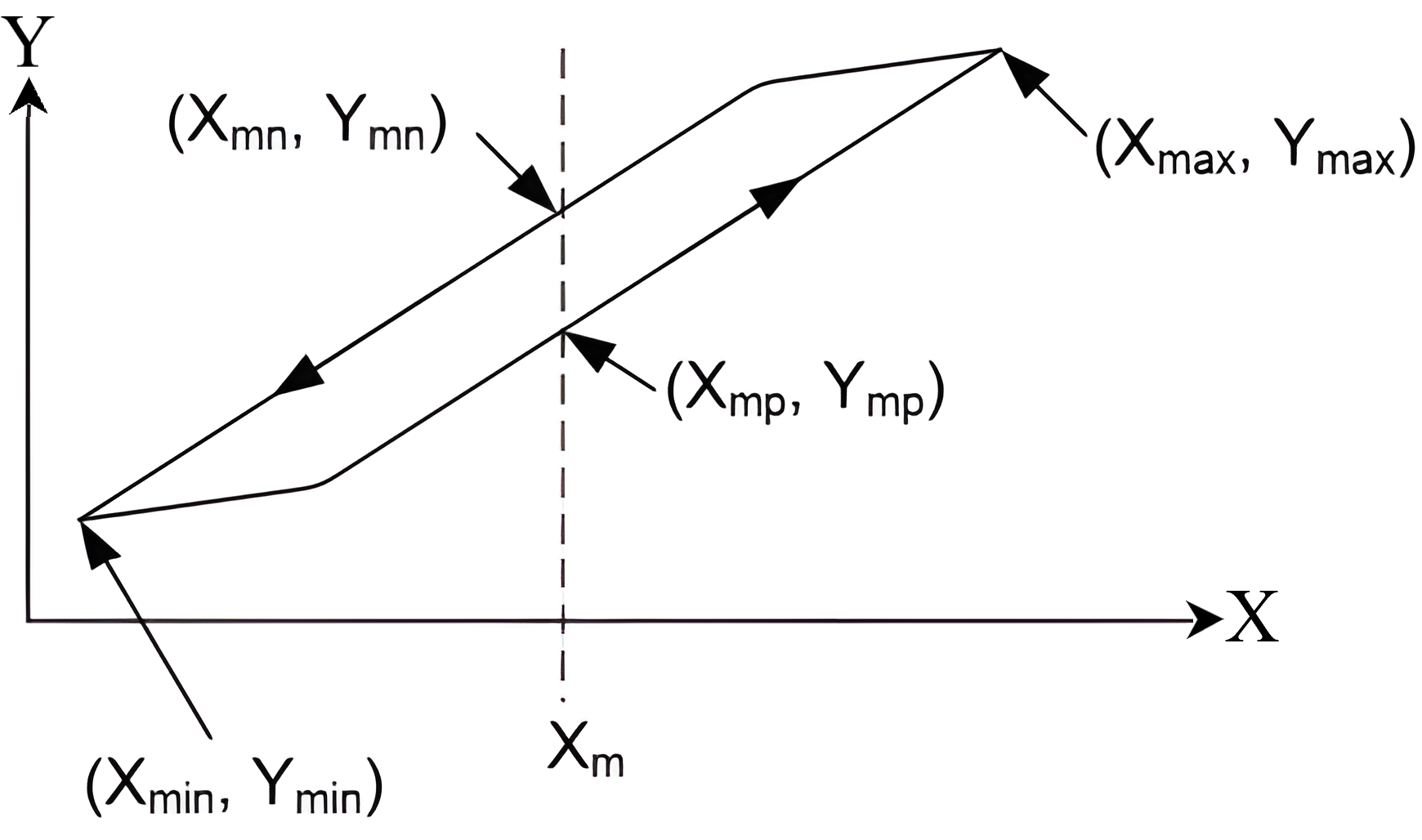}}
\caption{The procedure for identifying the required parameters to calculate the hysteresis error.}
\label{fig1}
\end{figure}

Fig. 13, shows the measured hysteresis curve for the proposed oscillator at 25 °C. The red (blue) curve shows the results for the case with increasing (decreasing) the water level. The main source of this error is the residual liquid film that remains on the sensor probe after the liquid level has been lowered. As can be seen, both cases are very similar and this confirms that the relative hysteresis error is minor and ignorable (it is less than 1.3\%).

Fig. 14, represents the measured output frequency versus changes in the surface level of the IV-fluid at a temperature of 25 °C. The RMNLE in this measurement is around 1.1\% at 25 °C. Figs. 11 to 14 provide insight into the sensor's sensitivity, nonlinearity, and repeatability. These results are valuable for understanding the behavior of the sensor under different conditions and highlight its great potential for different applications.
\begin{figure}[!h]
\centerline{\includegraphics[width=3 in]{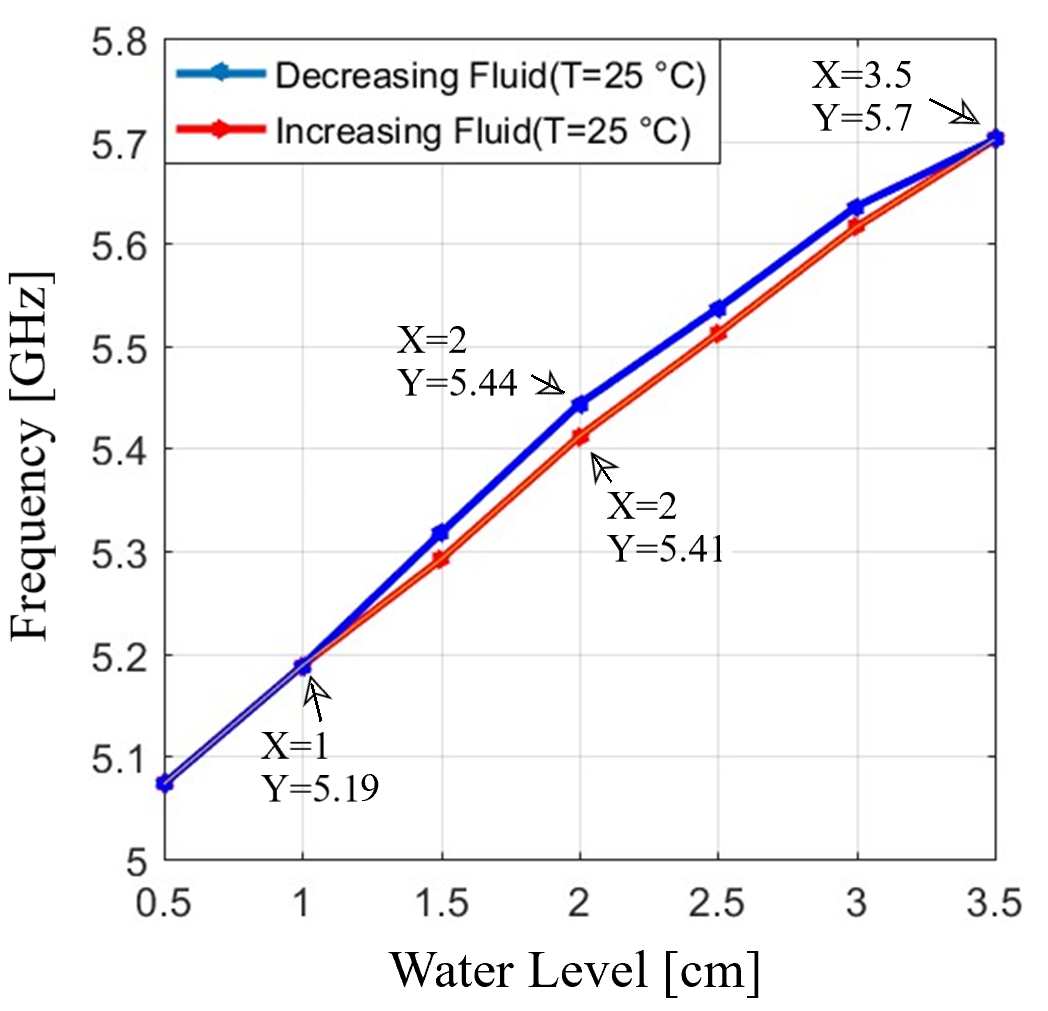}}
\caption{Output frequency versus water level changes at a temperature of 25 °C (upward and downward).}
\label{fig1}
\end{figure}
\begin{figure}[!h]
\centerline{\includegraphics[width=3 in]{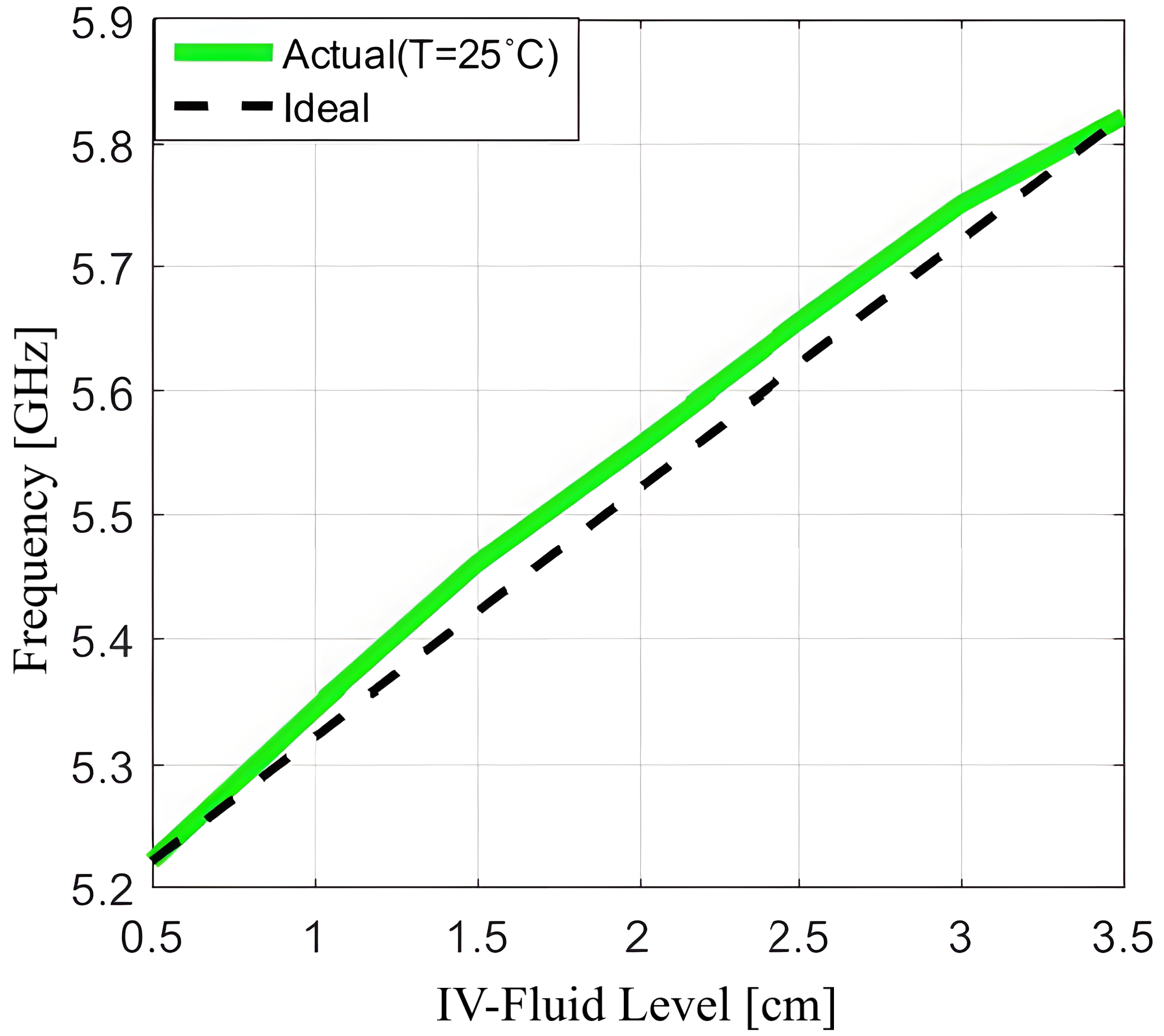}}
\caption{The measured output frequency versus IV-Fluid level changes.}
\label{fig1}
\end{figure}
To the best of our knowledge, no microwave oscillator level sensor has been reported so far. Therefore, Table III compares the sensor developed in this study with similar sensors from previous publications. The compared parameters include circuit dimensions, operating frequency, measurement range, sensitivity, LoD, RMNLE, and RANLE. Our sensor shows a superior performance in all aspects of sensitivity, LoD and linearity. Furthermore, it shows very robust behavior at different liquid temperatures. 
Radar and capacitive sensors are often used to measure the level of liquids. However, since their functionality is completely different from that of our sensor, a fair comparison is not possible, which is why they are not mentioned in this table. In addition, radar sensors are very expensive and require complex installation and calibration, and capacitor level sensors are highly sensitive to parasitic effects \cite{b11}. 

\begin{table}
\caption{Comparison with recently developed similar sensors.}
\label{table}
\label{tab1}
\centerline{\includegraphics[width=\columnwidth]{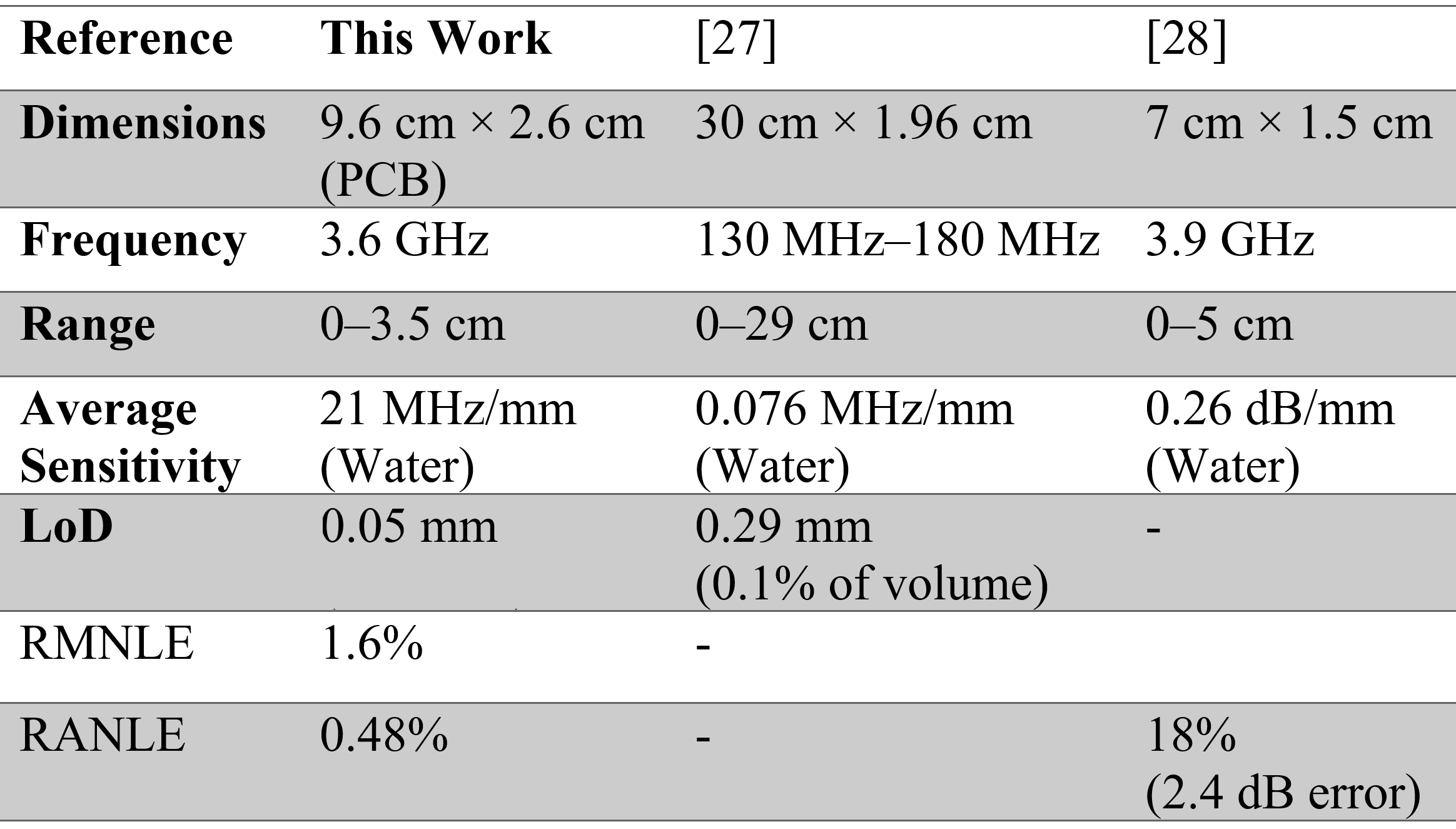}}
\end{table}

 \section*{Conclusion}

In the modern era, the utilization of wireless sensor networks has become increasingly prevalent for the measurement of various parameters of materials. This study introduces an oscillator sensor that offers an ideal combination of cost efficiency, sensitivity, and linearity. Such sensors find application not only in laboratory research but also in the expansive domain of wireless sensor networks (WSN).
One of the standout features of this sensor is its ability to adapt itself to the liquid permittivity, making it a valuable tool in the hands of researchers and engineers. It enables a precise adjustment of input resistance, a crucial aspect to ensure measurement stability and reliability. By fine-tuning the resistance value, users can attain the most linear and sensitive measurements possible for the liquid under test. This adaptability is particularly advantageous when dealing with materials with different dielectric constants, as illustrated in Fig. 6. The sensor range can be optimized to provide accurate results across different liquids.
This study substantiates its findings through numerical simulations using the CST 3D full-wave electromagnetic simulator combined with the Advanced Design System (ADS) nonlinear circuit simulations and practical experimentation. The good agreement between the simulation results and the experimental results underscores the robustness and efficacy of the developed sensor.
Furthermore, the study extends its investigation to liquid temperature variations, demonstrating the sensor's unwavering precision under different thermal conditions.

\section*{Acknowledgment}

The authors thank all members of the MEND (Modern Electromagnetics and Nanoelectronic Devices) research group at Constructor University for their valuable scientific and technical support.

\end{document}